\documentclass[%
 reprint,
 superscriptaddress,
 amsmath,amssymb,
 aps,
floatfix,
showkeys
]{revtex4-2}

\usepackage{xcolor}
\usepackage{graphicx}
\usepackage{dcolumn}
\usepackage{bm}
\usepackage{hyperref}
\usepackage{siunitx} 
\usepackage[mathlines]{lineno}
\linenumbers\relax
\usepackage{soul}
\usepackage{rotating}
\usepackage{subcaption}
\usepackage{url}
\usepackage{comment}
\usepackage[justification=raggedright,singlelinecheck=false]{caption}

\nolinenumbers

\begin{document}

\preprint{APS/123-QED}
\newcommand{\YS}[1]{\textbf{\textcolor{red}{YS: #1}}}

\title{Cosmic Axions Revealed via Amplified Modulation of Ellipticity of Laser (CARAMEL)
}

\author{Hooman Davoudiasl}
\affiliation{High Energy Theory Group, Physics Department, Brookhaven National Laboratory, Upton, New York 11973, USA}

\author{Yannis K. Semertzidis}
\thanks{Corresponding author}
\email{yannis@bnl.gov}
\affiliation{Department of Physics, Korea Advanced Institute of Science and Technology (KAIST), Daejeon 34141, Republic of Korea}

\begin{abstract}
We propose a new axion dark matter  detection strategy that employs optical readout of laser beam ellipticity modulations caused by axion-induced electric fields in a microwave cavity, using electro-optic (EO) crystals, enhanced by externally injected radio-frequency (RF) power. Building upon the variance-based probing method~\cite{Omarov_2023}, we extend this concept to the optical domain: a weak probe laser interacts with an EO crystal coupled to the resonant microwave cavity field at cryogenic temperatures, and the axion-induced electric field is revealed through induced ellipticity. The injected RF signal coherently interferes with that of the axion field, amplifying the optical response and significantly improving sensitivity. While our EO-based method employs a Fabry–Pérot resonator, we do not require Michelson interferometers.  Our method hence  enables compact, high-frequency axion searches, across the 0.5-50\,GHz range. Operating at cryogenic temperatures not only suppresses thermal backgrounds but, critically, allows the probing method to mitigate the quantum noise. This approach offers a scalable path forward for axion detection over the $\sim (\text{few}-200$)~$\mu$eV mass range  -- covering the preferred parameter space for the post-inflationary Peccei-Quinn axion dark matter --  using compact, tunable systems.

\end{abstract}


\maketitle

\section{Introduction}

The identity of dark matter (DM)  remains one of the most compelling unsolved problems in fundamental physics. Among the theoretically favored candidates is the quantum chromodynamics (QCD) axion $a$ \cite{article:Weinberg78,article:Wilczek78}, associated with the Peccei-Quinn (PQ) mechanism \cite{Peccei:1977hh,Peccei:1977ur} originally introduced to resolve the strong CP problem, {\it i.e.} the puzzling smallness of the CP violating parameter in QCD.  The axion of the PQ framework emerges  as a natural cold dark matter candidate, with its abundance  determined by early-universe dynamics \cite{Preskill:1982cy,article:Sikivie83,Dine:1981rt}. For a recent article on the resilience of the strong CP problem, see Ref.~\cite{kaplan2025solvestrongcpproblem}.

The mass $m_a$ of the QCD axion DM  is expected to have a narrow viable range if the associated PQ symmetry is broken after inflation in the early Universe (see, for example, Ref.~\cite{Marsh:2015xka}).  Cosmological and lattice calculations suggest that post-inflationary scenarios prefer an axion mass in the tens to hundreds of $\mu\mathrm{eV}$ \cite{Berkowitz:2015aua,Ballesteros:2016euj,Borsanyi:2016ksw,Dine:2017swf,article:post_infl2,Buschmann:2019icd} range, with a recent analysis suggesting $40~\mu{\rm eV} < m_a < 180$~$\mu$eV \cite{Buschmann:2021sdq}.  This mass range can be well covered by resonant cavity technique proposed below, in the photon frequency range 0.5-50\,GHz, which will be the focus of our discussion.

The most sensitive experimental approach to detecting axions in the above  mass window is the haloscope technique, which exploits axion--photon conversion in a resonant microwave cavity permeated by a strong magnetic field, as was suggested first by Sikivie~\cite{article:Sikivie83}. However, haloscopes face a fundamental limitation: as the resonant frequency increases, the physical volume of the cavity---and hence the signal power---must decrease. For an axion signal at 25\,GHz, a natural  cavity volume is $\sim 1~\mathrm{cm}^3$, over $10^5\times$ smaller than the volumes used in the original ADMX configuration~\cite{ADMX:2018gho,article:ADMX-anal,ADMX:2021nhd,article:ADMX-2024,article:ADMX-2025,ADMX:2025vom}.  Covering the 0.5-50~GHz range at half the resolution matched to the axion coherence bandwidth, $\Delta\nu/\nu \sim 10^{-6}$, requires $\sim  10^7$ distinct frequency channels (the band width for axions is set by their small kinetic energy $\propto v^2$, where $v\sim 10^{-3}$ is the virial velocity in our galactic neighborhood). This scaling imposes a stringent limitation on scanning speed, primarily because conventional readout electronics contribute significant quantum or thermal noise, requiring long integration times per frequency channel.  For example, assuming a typical $\sim 100$~s integration time per bin, covering the above frequency range would take $\sim 30$~years.   By contrast, our proposed axion signal probing method, based on laser readout through a Fabry-Pérot interferometer and electro-optic modulation, effectively bypasses the limitations of traditional RF electronics. It enables sensitivity well beyond the quantum noise floor typically imposed by the detection system.  The  method proposed here can potentially speed up scanning rates by more than an order of magnitude, using available technology.

Current state-of-the-art experiments in this regime rely on ultra-low-noise amplifiers, quantum-limited detectors, and often large optical interferometers or complex cavity geometries. Yet, sensitivity improvements remain incremental, and the ability to scan efficiently across millions of channels with sufficient signal-to-noise ratio (SNR) remains out of reach. A fundamentally different approach is required to break through this bottleneck---one that enables enhanced sensitivity even in small volumes, with suppressed noise, and minimal reliance on ultra-complex cryogenic amplification chains.

\begin{figure*}[t]
  \centering
\includegraphics[width=\textwidth]{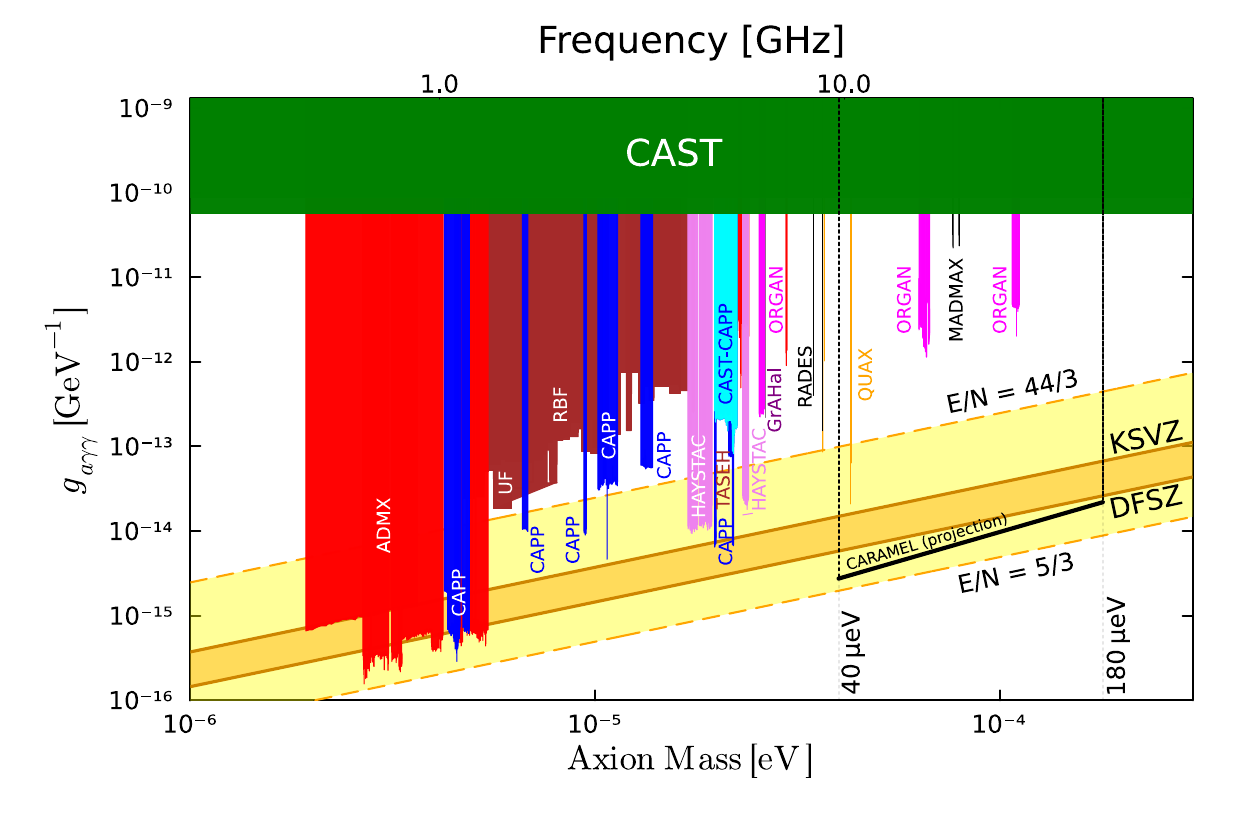}
\caption{The current status of the axion to two photon coupling in the mass range 1-300~$\mu$eV, corresponding to the frequency range 0.2-70\,GHz (from Ref.~\cite{AxionLimits}, ADMX~\cite{Asztalos2010,ADMX:2018gho,ADMX:2019uok,ADMX:2021nhd,Bartram:2024ovw,ADMX:2025vom}, ADMX-Sidecar~\cite{ADMX:2018ogs,Bartram:2021ysp}, ADMX-SLIC~\cite{Crisosto:2019fcj}, CAPP~\cite{Lee:2020cfj,Jeong:2020cwz,CAPP:2020utb,Yoon:2022gzp,Lee:2022mnc,article:CAPP-PACE-JPA,12TB-PRL,Yang:2023yry,Kim:2022hmg,article:Younggeun_Kim2024,article:Ahn-PRX2023,Bae:2024kmy}, CAST~\cite{CAST:2007jps,CAST:2017uph,CAST:2024eil}, CAST-CAPP~\cite{Adair:2022rtw}, GrAHal~\cite{Grenet:2021vbb}, HAYSTAC~\cite{Brubaker:2016ktl,HAYSTAC:2018rwy,HAYSTAC:2020kwv,HAYSTAC:2023cam,HAYSTAC:2024jch}, MADMAX~\cite{Garcia:2024xzc}, ORGAN~\cite{McAllister:2017lkb,Quiskamp:2022pks,Quiskamp:2023ehr,Quiskamp:2024oet}, QUAX~\cite{Alesini:2019ajt,Alesini:2020vny,Alesini:2022lnp,QUAX:2023gop,QUAX:2024fut}, RADES~\cite{CAST:2020rlf,Ahyoune:2024klt}, RBF~\cite{Panfilis1987,Wuensch:1989sa}, TASEH~\cite{TASEH:2022vvu}, UF~\cite{Hagmann,Hagmann:1996qd}). CARAMEL aims to facilitate  probing the frequency range of 0.5-50\,GHz, or $\sim$ (2-200)~$\mu$eV, with better than DFSZ sensitivity within the next five years.  
CARAMEL can potentially cover the preferred post-inflationary axion parameter space, corresponding to 40-180~$\mu$eV~\cite{Buschmann:2021sdq}, with better than DFSZ sensitivity using presently available technical capabilities. The projected sensitivity curve in this mass range corresponds to a total scanning time of approximately $2\times10^{7}$ s, which includes both data acquisition (3\,s) and cavity retuning (3\,s) per frequency step, assuming frequency steps of half the axion linewidth. The projection further assumes a constant axion-to-photon conversion power of $10^{-21}$\,W and a cavity quality factor of $Q_c=10^6$ across the indicated mass range (equivalent to $10^{-23}$\,W for a quality factor of $Q_c=10^4$).}
\label{fig:Axion_limits}
\end{figure*}

The current status of axion searches is summarized by the plot in Fig.\ref{fig:Axion_limits}, taken from Ref.~\cite{AxionLimits}.  As can be seen here, the window corresponding to the preferred post-inflationary PQ axion DM, $\sim 40-180$~$\mu$eV, within the KSVZ~\cite{Kim:1979if,Shifman:1979if} and DFSZ~\cite{Dine:1981rt,Zhitnitsky:1980tq} scenarios, remains largely unexplored.  This hints at the inherent challenges associated with this regime of parameters.   
 Progress in this range has been slow due to the convergence of several technical challenges: the cavity volume decreases with increasing frequency, conventional copper cavities suffer from degraded quality factors at high frequencies, and quantum-limited linear amplifiers constrain the readout sensitivity. Meanwhile, single-photon detectors, though promising, are not yet mature for robust operation under realistic laboratory conditions.

To overcome these barriers, multiple high-frequency haloscope concepts are under active development.  These include the pizza-cavity design~\cite{article:pizza_cavity}, which partitions the cavity into multiple sectors to increase frequency while preserving volume efficiency; photonic crystal~\cite{article:photonic_crystal,article:Bae2024} and metamaterial cavities~\cite{article:wheel_mechanism}, which decouple frequency from physical size; and horn-array haloscopes~\cite{article:Horn_2023}, which leverage large metal surface areas to achieve broadband, volume-efficient searches at high frequencies. Additionally, pioneering experiments such as MADMAX~\cite{article:MADMAX,article:MADMAX-2,article:MADMAX-3}, and DALI~\cite{article:DALI2025}  explore dielectric haloscope techniques using layered dielectric interfaces to coherently boost the axion-induced signal power. The ORGAN experiment~\cite{ORGAN2022,Organ2024,Organ2025} utilizes tunable resonant cavities in high magnetic fields to target axions in the 15–50\,GHz range with ultra-low-noise quantum-limited readout systems.
ALPHA uses the electron plasma resonance frequency of metalic rods~\cite{article:ALPHA2023}, and CAST-CAPP~\cite{article:CAST-CAPP} employed a decommissioned dipole magnet in an efficient geometry. Superconducting cavities with quality factors comparable to or exceeding the expected axion quality factor, even in the presence of strong magnetic fields (\(\sim 8\,\mathrm{T}\) and above), have been demonstrated using high-temperature superconducting (HTS) tapes~\cite{Danho19_1, Danho19_2, Danho22}. These developments indicate that this technology is maturing and offers a scalable path toward detecting axions in the tens to hundreds of \(\mu\mathrm{eV}\) mass range, as motivated by post-inflation QCD axion models.

In parallel, we introduce CARAMEL (Cosmic Axions Revealed via Amplified Modulation of Ellipticity of Laser), a novel detection technique that mitigates the effects of quantum noise and replaces conventional amplifier chains with optical readout. CARAMEL detects the axion-induced microwave field via its modulation of a laser beam  polarization or phase, thus bypassing the quantum noise limitations of linear amplification. Importantly, this method is not limited to high frequencies: it can also be applied at lower frequencies (e.g., near 200\,MHz), where electronic noise in standard amplifiers is still far above the quantum limit. In these cases, CARAMEL can yield sensitivity improvements of up to several orders of magnitude, making it a transformative technique for axion detection across the full spectrum from sub-GHz to tens of GHz.

Our proposal combines electro-optic (EO) detection~\cite{Ebadi:2023gne} and a ``resonant probing" technique~\cite{Omarov_2023}; see also~\cite{Riek_2015,Riek:2017vacuum,sikivie2021cleanenergydarkmatter}. These methods, previously suggested independently, are now unified to offer:
\begin{enumerate}
    \item Sensitivity mitigating the quantum noise limit at low temperatures, and
    \item Operation with only modest laser power (10\,mW).
\end{enumerate}
This approach is also applicable to other experimental geometries, including MADMAX and related dielectric haloscope designs~\cite{article:MADMAX,article:MADMAX-2,article:MADMAX-3,article:ALPHA2023,article:DALI2025}.

Next, we will discuss the basics of the axion detection in our proposed setup.  We will provide estimates for the signal and main sources of noise.  Our analysis will  illustrate that the SNR achieved can lead to a definitive detection across and beyond the  post-inflationary PQ axion DM parameter space, employing technology that is currently available.  
Henceforth, we shall assume a laser with power $P = 10\,\mathrm{mW} = 10^{-2}\,\mathrm{W}$ and wavelength  $\lambda = 1064\,\mathrm{nm} \Rightarrow \nu = c/\lambda \approx 2.82 \times 10^{14}\,\mathrm{Hz}$, where $\nu$ is the laser beam frequency; $h = 6.626 \times 10^{-34}\,\mathrm{J\, s}$ is Planck's constant.  

\begin{table}[b]
\caption{Benchmark parameters assumed in this work for the representative regime
$\nu_a \sim \mathcal{O}(10\,\mathrm{GHz})$. These parameters are employed throughout
the main text as a reference operating point to illustrate the scaling and
performance of the proposed method and are not unique or restrictive. A more
general and experimentally flexible parameter range, consistent with all
system-level requirements—including optical absorption, RF dissipation, and
cryogenic heat-loading constraints—is presented in Appendix~\ref{app:D} and summarized in
Table~\ref{tab:scaling}.}
\centering
\begin{tabular}{|l|c|}
\hline
Parameter & Value \\
\hline
Laser Power & 10 mW \\
Laser Wavelength & 1064 nm \\
RF Probe Power & 2 nW\\
Microwave Cavity Quality Factor $Q_c$  & $10^4$ \\
Axion-to-Photon Reference Power $P_a$, & \\
for $Q_c=10^4$& $10^{-23}$ W  \\
Microwave Cavity Volume & 3.7 L \\
Fabry-Pérot Finesse ${\cal F}$ & $10^4$ \\
EO Crystal (LiNbO$_3$) Thickness $L_c$ & 3 mm \\
\hline
\end{tabular}
\label{tab:benchmark}
\end{table}

\section{Detection Principle}

The basis of our proposal is the use of the EO effect, where an electric field $E\equiv |\vec{E}|$  leads to induced ellipticity $\psi$ in a laser beam of wavelength $\lambda$ (Pockels effect~\cite{yariv2006photonics,saleh2007fundamentals}).  We have  
\begin{equation}
    \psi = \frac{\pi}{\lambda} n^3 r_{ij}\, E \, L_c\,,
\label{eq:psi}
\end{equation}
where $n$ is refractive index of the EO crystal, $r_{ij}$ is assumed to be the largest element of the  EO coefficient tensor, and $L_c$ is the length of the crystal through which the laser propagates.  

Let us now estimate the contribution of axion conversion in a cavity to the above effect.  The power output from axion-to-photon conversion in a microwave cavity operating in the TM$_{010}$ mode, using experimental parameters as reported by CAPP, is given by~\cite{article:Ahn-PRX2023,Kim_2020}:  
\begin{widetext}
\begin{equation}
    P_{a\gamma\gamma} = 8.7 \times 10^{-23}\,{\rm W} \left(\frac{g_{\gamma}}{0.36}\right)^2 
    \left(\frac{\rho_a}{0.45\, {\rm GeV/cm^3}}\right) 
    \left(\frac{\nu_a}{1.1\,{\rm GHz}}\right) 
    \left(\frac{\langle \mathbf{B}_{e}^{2} \rangle}{(10.3\,{\rm T})^2}\right)
    \left(\frac{V}{37\,{\rm L}}\right) 
    \left(\frac{G}{0.6}\right) 
    \left(\frac{10^6}{Q_a}+\frac{10^5}{Q_c}\right)^{-1},
    \label{eq:conv_power}
\end{equation}
\end{widetext}
where $g_{a\gamma\gamma} = (\alpha/\pi)(g_{\gamma}/f_a)$ is the axion-photon coupling, $\alpha$ is the fine structure constant, and $g_{\gamma}$ takes values of $-0.97$ (KSVZ model \cite{Kim:1979if,Shifman:1979if}) or $0.36$ (DFSZ model \cite{Dine:1981rt,Zhitnitsky:1980tq}). $f_a$ is the PQ symmetry breaking scale,
also known as the decay constant. $\rho_a$ is the local axion dark matter density, and $\nu_a = m_a c^2/h$ is the axion Compton frequency ($c = 2.998 \times 10^8$ m/s is the speed of light).  In the above,  $\langle \mathbf{B}_{e}^2 \rangle$ is the squared average external magnetic field, $V$ is the cavity volume, $Q_a \approx 10^6$ is the axion quality factor, and $Q_c$ is the cavity unloaded quality factor. The form factor $G$ reflects the mode-dependent overlap of the axion-induced field with the resonant mode and is approximately 0.69 for the TM$_{010}$ mode.

In the limit $Q_c \ll Q_a$, the signal power scales linearly with $Q_c$:
\begin{equation}
    P_a \propto Q_c.
\end{equation}

Conversely, for $Q_c \gg Q_a$, the signal power saturates:
\begin{equation}
    P_a \approx \text{(constant)} \propto Q_a,
\end{equation}
reflecting the finite spectral width of the axion signal; the cavity cannot enhance the signal beyond the axion's intrinsic coherence bandwidth.

The total power deposited by axion conversion is given by
\begin{equation}
    P_a = \frac{\omega U}{Q_c},
\end{equation}
where $P_a$ is the axion-induced power in Eq.~(\ref{eq:conv_power}), $\omega = 2\pi \nu$ is the angular frequency of the cavity (or the axion, on resonance), and $U$ is the total stored energy in the cavity.  Since the stored energy is equally partitioned between electric and magnetic components, the electric field energy is~\cite{KIM2019100362} 
\begin{equation}
    U_E = \frac{1}{2} \epsilon_0 V E_a^2 = \frac{U}{2} = \frac{P_a Q_c}{2\omega},
\end{equation}
where $E_a$ is the root-mean-square electric field, $\epsilon_0 = 8.85 \times 10^{-12}\,\mathrm{F/m}$ is the vacuum permittivity, and $V$ is the cavity volume.  Solving for $E_a$ we get 
\begin{equation}
    E_a = \sqrt{ \frac{P_a Q_c}{\omega \epsilon_0 V} }\,.
    \label{eq:axionEfield}
\end{equation}
We will use $V = 3.7 \times 10^{-3}\,\mathrm{m^3}$ 
 (3.7 liters) as a benchmark value in what follows, appropriate for a modified 10\,GHz cavity. However, this volume is not a fixed requirement. At a given frequency, the cavity geometry may be adapted to vary $V$ while maintaining resonance to satisfy a desired reference power level. This flexibility in cavity volume is particularly useful for optimizing sensitivity while staying within the physical constraints imposed by the detection scheme and available magnet apertures.

We will consider using lithium niobate, LiNbO$_3$, of size $L_c=3$~mm, as our EO crystal, for which $n\approx 2.2$ and $r_{33}\approx 3.1\times 10^{-11}$m/V.  Let us take $\nu=10$~GHz as a representative value, with all the benchmark parameters used in this document shown in Table~\ref{tab:benchmark}.  We will consider the  benchmark values $P_a = 10^{-23}$~W and $Q_c=10^4$ for the chosen value of $\nu$ above.  From Eqs.(\ref{eq:psi}) and (\ref{eq:axionEfield}), we get $E_a \approx 7.0\times 10^{-9}$~V/m and $\psi_a \approx 2\times 10^{-14}$~rad. 

To determine the minimum detectable ellipticity, we need to estimate contributions from various sources of noise.  We identify two main sources:  (i) shot noise from the finite power output of the laser beam, and (ii) the quantum-thermal noise in the cavity.  Here, we first examine the effect of shot noise and in the next sub-section we will discuss quantum-thermal effects.  

{\it (i) Shot noise:} To estimate the photon shot noise, we use the standard expression
\begin{equation}
    \delta \psi_{\text{shot-laser}} = \frac{\sqrt{2}}{\sqrt{P/h\nu}} =  \sqrt{\frac{2 h\nu}{P}}
\end{equation}
where $P$ is the laser  power and $\nu$ is the associated frequency.  Substituting our benchmark values we have 
\begin{equation}
    \delta \psi_{\text{shot-laser }}   \approx 6.1 \times 10^{-9}\,\mathrm{rad \cdot Hz^{-1/2}}
    \label{eq:psi-shot}
\end{equation}

For an integration time of $t = 3$\,seconds, assumed henceforth, we get; see Appendix~\ref{app:A}:
\begin{equation}
    \delta \psi_{\text{shot-laser}} = \frac{6.1 \times 10^{-9}}{\sqrt{3}} \approx 3.5 \times 10^{-9}\,\mathrm{rad}\,.
\label{eq:delta-shot}
\end{equation}

Assuming only laser beam shot noise, we see that the expected typical SNR would be
\begin{equation}
\text{SNR}_{\rm laser} = \frac{\psi_a}{\delta \psi_{\text{shot}}}\approx 6\times 10^{-6}\,,
\label{axion-SNR}
\end{equation}
which is quite tiny.  

If we consider detection using only the axion-induced electric field $E_a$, a Fabry-Pérot (FP) cavity with finesse $\mathcal{F} = 10^4$ can boost the signal according to 
\begin{equation}
\psi_a \to  \left(\frac{2 \mathcal{F}}{\pi}\right) \psi_a \approx 1.3 \times 10^{-10} \, \text{rad}\,, 
\end{equation}
that would yield an $\text{SNR}_{\rm FP}\sim  0.04$, which is insufficient for detection.  However, we will see next how using a probe radio-frequency (RF)  beam at the axion frequency of oscillation can further enhance the signal.  

\subsection{Probing Method with Electro-Optic Readout}

The axion-induced oscillating electric field modulates the birefringence of the EO crystal, imprinting a small ellipticity or phase shift onto the laser polarization. To enhance the readout signal, we introduce an RF tone at the cavity resonance frequency, which coherently interferes with the axion-induced field inside the cavity. The probing method effectively amplifies the signal by creating a beat between the injected RF field and the axion signal, leading to a variance in the optical response that is detectable with high sensitivity. This technique, originally developed for microwave detection using power detectors~\cite{Omarov_2023}, is applied in the EO axion signal detection here, for the first time.

Since the detection occurs via polarization rotation (ellipticity) in the optical field---not via amplification of the microwave field itself---the signal readout is not constrained by the standard quantum limit. Furthermore, because the probing method effectively decouples the signal strength from the laser power, the system can operate at low optical intensities, minimizing thermal loading on the cryogenic environment.

Operating at cryogenic temperatures ($T \lesssim 100$\,mK), the thermal photon background for a single photon detector is exponentially suppressed. However, quantum field fluctuations still exist for the method suggested here, albeit significantly reduced. Combined with the narrow bandwidth of a high-$Q_c$ cavity, this results in a low-noise environment ideal for detecting weak axion-induced fields. The result is a highly sensitive, compact, and scalable architecture for axion detection in the 0.5--50\,GHz range; in principle the lower range can be further reduced.

Next, we will examine implementing the probing method to provide a feasible detection approach, without FP enhancement.  We will consider an injected RF power $P = 2\,\mathrm{nW} = 2 \times 10^{-9}\,\mathrm{W}$. The electric field amplitude, for $\omega=2 \pi \times 10^{10}$ rad/s, $V=3.7\,l$, and $Q_c = 10^4$ is:

\begin{equation}
    E_{\text{probe}} = \sqrt{\frac{Q_c P}{\epsilon_0 V \omega}} \approx 0.1 \,\mathrm{V/m}
\end{equation}

In the probing method, we are interested in detecting the fluctuations of the ellipticity, which are enhanced by the RF injected power.  To see this, we note that the total electric field in the cavity is given by  

\begin{equation}
    E(t) = E_{\text{probe}} \cos(\omega t) + E_a \cos(\omega t + \phi(t)),
\end{equation}
where $\phi(t)$ is the time-dependent relative phase between the two fields.

The axion field is not perfectly monochromatic; it has a finite spectral width $\Delta \nu_a \sim 10^{-6} \nu$ due to the virialized velocity dispersion of dark matter in the galactic halo. At $\nu = 10 \, \mathrm{GHz}$, this corresponds to $\Delta \nu_a \sim \mathrm{10\, kHz}$ and a coherence time of
\begin{equation}
    \tau_a \sim \frac{1}{\Delta \nu_a} \sim 0.1 \, \mathrm{ms}.
\end{equation}
Over timescales shorter than $\tau_a$, the axion field behaves like a coherent wave with a well-defined phase. Over longer timescales, this phase $\phi(t)$ drifts randomly, reflecting the stochastic nature of the axion field.

In the probing method, one is interested in measuring the variance of the ellipticity 
\begin{equation}
    \langle \psi^2 \rangle \propto \frac{1}{2}\left[E_{\text{probe}}^2 + E_a^2 + 2 E_{\text{probe}} E_a \langle \cos(\phi(t)) 
    \rangle\right],
    \label{eq:psisq}
\end{equation}
where $\langle \cos(\phi(t)) \rangle$ vanishes over many coherence times, but the variance of the fluctuations is nonzero.  To capture these fluctuations, one needs to have a sampling time scale $t_{\rm sample} \lesssim \tau_a$.  In the above expression (\ref{eq:psisq}), fast axion oscillations at $\omega \sim m_a$ have been averaged out, since they take place at time scales much shorter than $t_{\rm sample}$. Physically, the probe tone mixes coherently with the axion-induced field to generate a beat signal at the difference frequency — typically in the 1–50\,kHz range. This beat modulates the birefringence of the electro-optic crystal at a frequency far below the carrier, allowing low-frequency optical detection of an otherwise extremely weak GHz signal.

Given the above discussion, the detected ellipticity scales with the geometric mean of the axion and probe electric fields:
\begin{equation}
    \psi_{\rm probe} = \frac{\pi}{\lambda} n^3 r_{33} \sqrt{E_a E_{\text{probe}}}\, L_c.
\end{equation}
Substituting the above benchmark values, we find $\psi_{\rm probe} \approx 7.7 \times 10^{-11}$~rad.  As before, if we employ a FP cavity with a finesse ${\cal F}$, we can enhance the signal 
\begin{equation}
\psi_{\rm probe} \to \left(\frac{2 \mathcal{F}}{\pi}\right) \psi_{\rm probe}\,,
\label{eq:FP-psidet}
\end{equation}
which for ${\cal F}=10^4$ will result in $\psi_{\rm probe} \approx 4.9
\times 10^{-7}$~rad.

Thus, the SNR for a 10\,mW laser power and integrating for 3\,s, is :
\begin{equation}
    \mathrm{SNR}_{\rm probe-FP} \approx \frac{(2 {\cal F}/\pi) \psi_{\rm probe}}{\delta\psi_{\text{shot}}} \approx \frac{4.9\times 10^{-7}}{3.5  \times 10^{-9}} \approx  140
\end{equation}
We see, form the above, that a combination of probing and FP enhancement allows for a detectable signal, easily overcoming the measurement limitations. In fact, given the size of the above SNR, we could envision that a lower laser beam power, in conjunction with a thinner EO crystal, would also suffice for the measurement.

Here, we would like to make the following two points.  First, as mentioned earlier, due to the probing method, the required  sampling time is set by the axion coherence time  $\tau_a=Q_a/\nu$, which is $\sim 10$\,kHz, about 6 orders of magnitude below the axion frequency.  Secondly, the RF probe affects both the axion signal and the quantum-thermal background.  This suggests that we need to examine the associated effective SNR for the measurement further, which will do next.   

\subsection{RF probe effect on sensitivity}

Vacuum fluctuations of the electromagnetic field, though classically absent, have measurable consequences in quantum electrodynamics. In particular, they give rise to the zero-point energy of each electromagnetic mode, with an average energy of \(\hbar \omega / 2\), and manifest as non-zero field quadratures even at zero temperature.  Single-photon detectors are sensitive only to real (on-shell) photons through energy absorption, and hence they remain insensitive to these vacuum fluctuations.  However, we would need to account for such off-shell effects, as discussed below.

{\it (ii) Quantum-thermal noise:} The average number of thermal photons in the cavity mode is:
\begin{equation}
\bar{n}_{\mathrm{th}} = \frac{1}{e^{\hbar \omega / k_B T} - 1} \,,
\end{equation}
where $k_B=1.38 \times 10^{-23}$\,J/K is the Boltzmann constant.  
Each photon contributes an energy $\hbar \omega$, with $\hbar \equiv h/(2 \pi)$, so the thermal energy associated with the cavity is:
\begin{equation}
U_{\mathrm{th}} =  \hbar \omega \left( \bar{n}_{\mathrm{th}} +\frac{1}{2}\right)
\end{equation}
with the half photon constant background present due to off-shell vacuum fluctuations.

This energy is equally partitioned between the electric and magnetic fields, so the electric field energy is
\begin{equation}
    U_E = \frac{1}{2} U_{\mathrm{th}}
\end{equation}
The above energy can also be expressed in terms of the electric field amplitude as
\begin{equation}
    U_E = \frac{1}{2} \epsilon_0 V E_{\text{th}}^2.
\end{equation}
Solving for the  electric field $E_{\text{th}}$ yields
\begin{equation}
E_{\mathrm{th}} = \sqrt{\frac{ U_{\mathrm{th}}}{\epsilon_0 V }} = \sqrt{\frac{ (\bar{n}_{\mathrm{th}}+1/2) \hbar \omega}{\epsilon_0 V }}
\end{equation}
which represents the root-mean-square electric field amplitude of the thermal plus quantum fluctuations in the cavity. This sets the quantum noise floor for field-based detection methods.  

The axion-induced power is:
\begin{equation}
    P_a = P_0\, Q_{\rm red} ,
\end{equation}
where $P_0$ is a reference quantity and the reduced quality factor given by 
\begin{equation}
Q_{\rm red}\equiv 
\frac{Q_c Q_a}{Q_c + Q_a}.
\label{eq:Qred}
\end{equation}
For our benchmark choices  $P_0 = 10^{-27}$~W, for $Q_c=10^4$.  The stored electric field amplitude in the resonant cavity is given by Eq.(\ref{eq:axionEfield}), as derived before 
\begin{equation}\nonumber
    E_a = \sqrt{ \frac{P_a Q_c}{\epsilon_0 V \omega} }.
\end{equation}

Assuming an axion coherence time $\tau_a = Q_a / \nu $, the signal-to-noise ratio (SNR) due to thermal noise after an integration time $ t $ is given by:
\begin{equation}
    \mathrm{SNR}_{\rm th}(t) = \frac{E_a}{E_{\mathrm{th}}} \cdot \sqrt{ \frac{t}{\tau_a} },
\end{equation}
which yields
\begin{equation}
    \mathrm{SNR}_{\rm th}(t) = 
    \sqrt{ \frac{P_0 Q_{\rm red} Q_c t }{2\pi(\bar{n}_{\mathrm{th}} + \tfrac{1}{2}) \hbar \omega Q_a} },
    \label{eq:SNRt}
\end{equation}
valid when $Q_c\le Q_a$.
Equation~(\ref{eq:SNRt}) treats the full cavity bandwidth as contributing thermal noise and
therefore provides only an approximate estimate.  
In practice, only the noise within the axion (or beat-note) linewidth is
relevant, introducing a bandwidth factor $\sqrt{Q_c/Q_a}$ that Eq.~(\ref{eq:SNRt}) does not
include.  
The complete treatment in Appendix~\ref{app:B}, (Eq.~\ref{Eq:general}) provides: 
\begin{equation}
\boxed{
\mathrm{SNR}(t) = \sqrt{\frac{ P_0 Q_{\text{red}}t}{ \pi \hbar \omega \coth(\frac{\hbar \omega}{2 k_B T})}}
}.
\label{Eq:main}
\end{equation}
This expression smoothly interpolates between the quantum regime at low temperatures [where $\coth (\hbar \omega/2 k_B T)\to 1$ as $T\to 0$] and the classical limit at high temperatures [where $\coth (\hbar \omega/2 k_B T) \to 2k_B T/\hbar \omega$ as $T\to \infty$].

\begin{figure}[t]
  \centering
    \includegraphics[width=\linewidth]{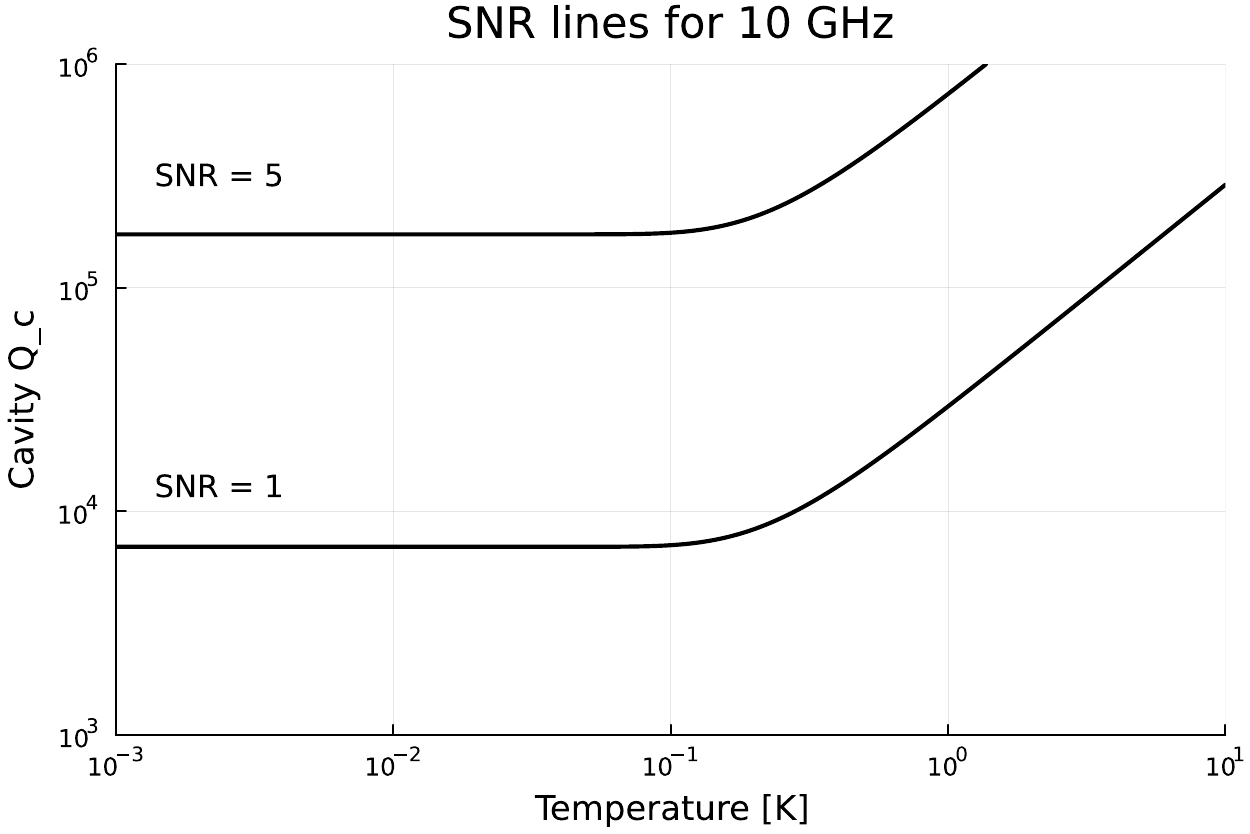}
    \caption{Signal-to-noise ratio   (SNR) as a function of $Q_c$ and $T$ at different operating frequencies. The transition from the quantum regime to the classical regime becomes apparent around 480\,mK,  where the thermal photon occupation number begins to exceed the vacuum (zero-point) contribution. The cavity volume is kept constant  at 3.7 liters. The axion to photon conversion power is kept at $10^{-23}$\,W  assuming $Q_c=10^4$, $t=3$\,s, and scaled appropriately for different cavity quality factor values; see Appendix B and D.}
  \label{fig:snr_plot10GHz}
\end{figure}

As detailed in the Appendix, the thermal noise field contributing to signal detection is not determined solely by the total thermal energy stored in the cavity mode but by the fraction of this energy that spectrally overlaps with the axion signal bandwidth. In the regime where the cavity quality factor is much smaller than the axion coherence quality factor ($Q_c \ll Q_a$), only thermal fluctuations within a bandwidth $\Delta \nu \sim \nu/Q_a$ contribute effectively to the signal readout. This leads to a suppression of the thermal noise electric field amplitude by a factor $\sqrt{Q_c/Q_a}$, since the axion signal is narrow band while the cavity noise is broadband. The probing method, by isolating this narrow interference bandwidth, enables a higher SNR despite a broad cavity response. Figure~\ref{fig:snr_plot10GHz} illustrates this effect for the representative case at 10\,GHz, showing how SNR depends on both temperature and cavity quality factor, and highlighting the transition between quantum and classical thermal noise regimes.

The $\sqrt{Q_c/Q_a}$ thermal noise suppression can also be recovered without probing, provided that the cavity output is analyzed with frequency resolution matched to the axion linewidth $\Delta \nu_a = \nu/Q_a$. In this case, only the thermal noise within a single high-resolution Fourier transform bin contributes, and the SNR expression remains equivalent. However,  the probing method achieves the same filtering effect automatically through beat-based detection.

\section{Discussion: Surpassing the Laser Quantum Noise Limit
 of Linear Microwave Amplification}
\label{section:discussion}

The combination of the probing method—where a known RF signal is injected into the system—and FP enhancement enables a substantial amplification of the signal. In particular, the induced ellipticity benefits from the geometric mean field $\sqrt{E_a E_{\text{probe}}}$, effectively boosting the axion signal strength. This hybrid approach allows the experiment to surpass the traditional quantum shot-noise limit associated with laser light. As shown in Fig.~2 of Ref.~\cite{Omarov_2023}, the injection of probe photons significantly enhances the signal-to-noise ratio (SNR) when the laser shot noise exceeds the sampling rate, corresponding to regions II and III in the figure. In contrast, when the shot noise is lower than the sampling rate (regions I and IV), the addition of probing power leads to reduced SNR due to unnecessary variance amplification in the electro-optic signal readout.

Owing to the resulting enhancement in signal-to-noise ratio (SNR)—by several orders of magnitude—this method offers considerable flexibility in experimental design. For example, one can reduce the laser power from 10 mW to 1 mW, while maintaining sensitivity to axions with DFSZ-level coupling.  Future developments could include spatially optimized electro-optic (EO) geometries and dynamic modulation of the probe signal to further suppress noise and improve detection fidelity.  Since the axion-probe beat signal appears at low frequencies (typically 1–50\,kHz), laser phase noise is not a limiting factor, provided a low-noise narrow-linewidth laser is used. Locking the laser to a Fabry–Pérot cavity further suppresses phase fluctuations in this band. While probing techniques can, in principle, introduce their own quantum limits, in the present optical readout scheme such effects are not expected to pose a significant limitation. The dominant quantum noise source remains the vacuum fluctuations of the resonant cavity field. Since the injected RF probe is a coherent classical signal and the detection process involves measuring optical ellipticity rather than amplifying the microwave field, any additional quantum noise introduced by probing is expected to lie below the vacuum noise floor. As such, the overall sensitivity is ultimately limited by the cavity's zero-point fluctuations, which are already accounted for in our analysis.

Until practical single-photon detectors become available in the GHz regime~\cite{article:SPC_QLA, article:SPD, article:Kuzmin2020, article:Gatti2023,article:SPC_QLA,article:Kuzmin2020,article:Mottonen2016,article:CARRACK, article:Ahn23,pankratov2025_CASH},
the method presented here offers a compelling enhancement in both sensitivity and scanning speed. Specifically, this approach can accelerate axion searches by more than an order of magnitude through two mechanisms. First, by injecting a coherent probe tone and detecting the axion-induced beat signal via an EO  crystal, the effective SNR  exceeds the noise limit of conventional haloscopes by a factor of 5--10~\cite{Panfilis1987, article:Hagmann1990,CAST:2017uph,ADMX:2018gho, ADMX:2019uok, ADMX:2021nhd,Brubaker:2016ktl, article:HAYSTAC-2, article:HAYSTAC-3,Lee:2020cfj, article:CAPP-PACE-JPA, Jeong:2020cwz, CAPP:2020utb, 12TB-PRL,article:CAST-CAPP,article:CAPP18T2023, Lee:2022mnc,article:Younggeun_Kim2024,grenet2021grenoble,article:Bae2024}. Since the scanning rate scales inversely with the square of the SNR, this corresponds to a 25--100$\times$ increase in scanning speed while still achieving DFSZ sensitivity, even with the modest integration time of 3 seconds per frequency step assumed throughout this work. Second, the same Fabry–Pérot laser readout system can be used uniformly across a wide frequency range—from 0.5\,GHz to at least 50\,GHz—without reconfiguring the detection chain. This broadband capability is a critical advantage over conventional linear amplifiers and quantum-limited photon counters, which typically require frequency-specific optimization. Together, these features make the present method a versatile and powerful strategy for exploring the high-frequency axion parameter space that remains largely open.
For more discussion on this subject, see  Appendix~\ref{app:C}.

\subsection*{Leveraging Axion Spatial Coherence}

An immediate and practical extension of the present method would be to deploy a modest array of probing haloscopes operating in parallel. Given that vacuum fluctuations are uncorrelated between separate cavities, while the axion field remains spatially coherent over macroscopic scales, coherently combining the outputs of $N$ detectors leads to a signal that scales as $N$, while the noise grows only as $\sqrt{N}$. Thus, the overall signal-to-noise ratio improves as
\begin{equation}
\mathrm{SNR}_{\text{total}} \propto \frac{N}{\sqrt{N}} = \sqrt{N} \quad,
\nonumber
\end{equation}
for fixed total integration time, equivalent to a single search with a volume $N V$.

This strategy is justified by the fact that the axion field is expected to behave as a classical, spatially coherent oscillation over a characteristic coherence length
\begin{equation}
\ell_{\text{coh}} \sim \frac{2\pi}{m_a v} \sim  \text{100 m to } 1\,\mathrm{km},
\end{equation}
depending on the axion mass $m_a$ (or frequency $\nu_a$) and its virial velocity $v \sim 10^{-3}$. For example, at $\nu_a = 1\,\mathrm{GHz}$, the coherence length is approximately $200\,\mathrm{m}$, ensuring that ten or more detectors within this radius observe the same axion phase and can be summed coherently.

The distinction between coherent axion signal and incoherent quantum or thermal noise is especially important when targeting higher-frequency axions. In that regime, the resonant cavity volume must shrink to maintain the desired frequency, limiting the available signal power. One can compensate for the loss of volume by launching several experiments in parallel, tuned at the same frequency. Since the optical Fabry--Pérot readout system is compact, stable, and easily replicated, a tenfold array is straightforward to implement with current technology and could provide the required improvement in sensitivity or scanning rate---without introducing additional quantum or thermal noise limitations.

\subsection*{Implications for Axion Haloscopes}

We note that the EO-based electric field readout method, especially when combined with probing, could offer significant advantages to other haloscope-style axion experiments, such as, e.g., ADMX, MADMAX and ALPHA.

In MADMAX, the dielectric disk configuration enhances axion-photon conversion but currently requires extremely low noise power detection. Incorporating EO-based field readout could allow detection of the same signal with reduced thermal and quantum noise concerns, potentially allowing operation at 1\,K or even higher.

The ALPHA experiment also operates at higher frequencies in compact resonators and faces similar challenges from quantum noise and dilution refrigeration complexity. Because EO readout responds to electric field amplitude rather than energy, it provides a more favorable SNR scaling and offers a path forward for simplified cryogenic operation. A key advantage of the proposed approach is the elimination of readout noise as a limiting factor in axion detection. By replacing conventional RF amplification chains with an optical probing system using a coherent laser and Fabry-Pérot interferometry, we suppress the dominant quantum and thermal noise sources that typically constrain integration times. This unlocks significantly faster scanning rates and ensures sensitivity to axion signals across a wide frequency range—independent of the electronic noise floor that currently limits haloscope performance.

Together, these implications suggest that EO-based probing methods, represent a compelling new detection paradigm for future axion searches.

\section{Concluding Remarks and Summary}

The CARAMEL method proposed in this work achieves a decisive advantage by replacing quantum-limited microwave amplification with a low-noise optical readout, even though the axion-photon conversion mechanism and the resonator design are conceptually identical to those used in traditional haloscope experiments.  This readout scheme leverages the EO effect to convert weak electric field modulations into laser polarization signals, which can be measured with shot-noise-limited sensitivity over narrow detection bandwidths. In doing so, CARAMEL retains all the standard haloscope principles while extending their reach through a fundamentally different detection chain.

In addition to its low-noise readout advantage, CARAMEL offers two notable system-level benefits. First, it enables a unified detection platform across a vast frequency range, from sub-GHz to tens of GHz, without the need for reconfiguring quantum-limited amplifiers or microwave components. Second, it eliminates the added noise introduced by microwave amplification, which—even at the quantum limit—contributes at least half a photon of noise power to the system. The optical readout scheme in CARAMEL bypasses this fundamental amplifier constraint, allowing high-sensitivity measurements without incurring quantum-limited amplification penalties.  We note that the CARAMEL approach does not beat vacuum noise in the absolute sense, but it resolves an axion signal in a narrow bin where vacuum noise is statistically small. This is achieved through the combined use of probing, Fabry--Pérot enhancement, and long integration.  Thus, our method does not eliminate vacuum fluctuations — it resolves the axion field against them using time and coherence. We emphasize that the EO/FP stage serves as an optical field amplifier, not a resonant element at the axion frequency. As a result, the axion search bandwidth is determined solely by the microwave cavity and the bandwidth of the optical detection chain, rather than by the Fabry–Pérot resonator.

In summary, CARAMEL unifies optical and RF techniques to achieve axion dark matter detection with enhanced sensitivity beyond the quantum noise limit. Its ability to operate across a broad frequency range, with low optical power and without complex quantum amplifiers, opens a scalable path forward in the high-frequency regime. Our approach is complementary to existing axion haloscopes and provides a versatile platform for future axion dark matter searches.  The method described in this work is projected to cover much -- perhaps all \cite{Buschmann:2021sdq} -- the relevant parameter space for post-inflationary axion dark matter, and may  lead to a breakthrough in resolving puzzles regarding both the sub-atomic interactions of hadrons and the large scale structure of the Universe.

\begin{acknowledgments}
The work of H.D. is supported by the US Department of Energy under Grant Contract DE-SC0012704. YKS is indebted to KAIST and in particular to the Physics Department for support to finish this work. YKS acknowledges fruitful discussions with Junu Jeong on the impact of zero-point fluctuations on the signal-to-noise ratio in single versus multiple cavity configurations.
\end{acknowledgments}

\newpage

\appendix

\section{Laser Signal-to-Noise Ratio Estimation}
\label{app:A}

We estimate the signal-to-noise ratio (SNR) for the axion-induced ellipticity in the probing method with optical readout.

The shot-noise-limited ellipticity sensitivity of the laser readout system is given by
\begin{equation}
    \delta \psi_{\text{shot}} =  \sqrt{ \frac{2h \nu}{ P } } \approx 6.1\,\text{nrad}/\sqrt{\text{Hz}},
\end{equation}
where   $\nu  \approx 2.82 \times 10^{14}\,\mathrm{Hz}$ is the laser frequency, \( P=10\, \mathrm {mW} \) the laser power. 

The axion-induced ellipticity is enhanced by the externally injected RF probing field and the finesse of the optical cavity. For the chosen parameters of FP finesse of $10^4$ and RF probing power of 2\,nW, the expected signal amplitude is
\begin{equation}
    \psi_{a} \approx 490\,\text{nrad}.
\end{equation}

The axion field has a finite coherence time of approximately
\begin{equation}
    \tau_a = \frac{1}{\Delta\nu_a} \approx 100\,\mu\text{s},
\end{equation}
corresponding to a bandwidth of \( \Delta\nu_a = 10\,\text{kHz} \). During a total integration time \( t = 3\,\text{s} \), the number of independent coherence segments is
\begin{equation}
    N = \frac{t}{\tau_a} = t \cdot \Delta\nu_a = 3\,\text{s} \times 10^4\,\text{Hz} = 3 \times 10^4.
\end{equation}

The shot noise within each coherence time is
\begin{equation}
    \delta\psi_{\text{seg}} = \delta\psi_{\text{shot}} \cdot \sqrt{\Delta\nu_a} = 6.1\,\text{nrad} \cdot 100 = 610\,\text{nrad}.
\end{equation}

Averaging incoherently over \( N \) segments, the total noise decreases as
\begin{equation}
    \delta\psi = \frac{610\,\text{nrad}}{\sqrt{3 \times 10^4}} \approx 3.5\,\text{nrad}.
\end{equation}

Therefore, the final SNR is given by
\begin{equation}
    \text{SNR} = \frac{\psi_a}{\delta\psi} = \frac{490\,\text{nrad}}{3.5\,\text{nrad}} \approx 140.
\end{equation}

This confirms that the probing method with optical readout can detect the axion-induced signal at high significance within a short integration time, provided the axion frequency is on resonance.

\section{Signal-to-Noise Ratio Dependence on Quality Factor and Temperature}
\label{app:B}

Here, we clarify the proper use of thermal noise field expressions and demonstrate how the probing method automatically reduces the thermal noise and improves the signal-to-noise ratio (SNR), especially when the cavity quality factor is lower than the axion coherence quality factor, $Q_c \ll Q_a$. The same effect can be achieved by effectively integrating over the axion band width.

\subsection*{B1. Thermal Noise in a Resonant Cavity Mode}

For a single resonant mode at angular frequency $\omega = 2\pi \nu$, the average thermal energy in the mode is:
\begin{equation}
U_{\text{mode}} = \left(n_{\text{th}} + \tfrac{1}{2}\right) \hbar \omega,
\end{equation}
where $n_{\text{th}} = 1/(e^{\hbar \omega / k_B T} - 1)$ is the thermal occupation number.

This energy is shared equally between the electric and magnetic fields. The electric field RMS amplitude is therefore:
\begin{equation}
E_{\text{th}}^{\text{(mode)}} = \sqrt{ \frac{(n_{\text{th}} + \tfrac{1}{2}) \hbar \omega }{ \epsilon_0 V } },
\end{equation}
where $V$ is the effective mode volume. In a more compact form $U_{\text{mode}}= \frac{\hbar \omega}{2} \coth  \left( \frac{\hbar \omega}{2 k_B T} \right ) $ and so,
\begin{equation}
E_{\text{th}}^{\text{(mode)}} = \sqrt{ \frac{ \hbar \omega \coth \left( \frac  {\hbar \omega}{2 k_B T} \right )}{2 \epsilon_0 V } },
\label{eq:thermE}
\end{equation}
which is valid in both the classical and quantum limit.

\subsection*{B2. Thermal Noise in a Band-Limited Measurement}

In the classical limit, the thermal noise has a flat power spectral density near GHz frequencies, given by:
\begin{equation}
dP_{\text{noise}}/d \nu = k_B T \quad \text{[W/Hz]}.
\end{equation}
When a detector samples a bandwidth $\Delta \nu$ for a time $\Delta t$, the total thermal energy collected is:
\begin{equation}
U = k_B T \cdot \Delta \nu \cdot \Delta t.
\end{equation}
Assuming half of this energy contributes  to the electric field and half to the magnetic field~\cite{KIM2019100362}, the corresponding RMS E-field amplitude is:
\begin{equation}
E_{\text{th}}^{\text{(probe)}} = \sqrt{ \frac{ k_B T \cdot \Delta \nu \cdot \Delta t }{ \epsilon_0 V } }.
\end{equation}
Let us introduce the reduced quality factor $Q_{\text{red}}= Q_a Q_c/(Q_a+Q_c)$, which we will use below.

\subsection*{B3. The Effect of Probing on SNR in the Low Quality Factor Regime}

In the regime $Q_c \ll Q_a$, the axion signal is spectrally narrow ($\Delta \nu_a = \omega/2\pi Q_a$), but the cavity bandwidth $\Delta \nu_c = \omega/2\pi Q_c$ is broad. This mismatch would ordinarily result in collecting excess thermal noise. 

However, the probing method changes this: a narrowband probe tone is injected at frequency $\omega_p$ near the axion frequency $\omega_a$, and the detection system is sensitive only to the interference between the axion field and the probe. This effectively limits the measurement bandwidth to $\Delta \nu = \omega / (2\pi Q_a)$, matching the axion coherence bandwidth. Since the axion field is coherent over a time $\tau_a  = Q_a /  \nu$ (more accurately, $\tau_{\text {coherence}}  = (Q_a +Q_c) /  \nu$, but we will use the above definition for simplicity when $Q_c \ll Q_a$), the measurement remains Fourier-limited with $\Delta \nu \cdot \tau_a = 1$. The same effect is obtained by integrating the signal only over the axion band width. If this is not possible, the probing can be used to achieve the narrow filtering automatically.

Although the cavity supports thermal noise across its full bandwidth, only thermal fluctuations within the narrow axion bandwidth contribute to the detected beat signal. This acts as an effective filtering mechanism, and the resulting thermal noise field is reduced by a factor $\sqrt{Q_c/Q_a}$ compared to the full cavity thermal field:
\begin{equation}
E_{\text{th}}^{\text{(probe)}} = \sqrt{ \frac{k_B T}{\epsilon_0 V} } \cdot \sqrt{ \frac{Q_c}{Q_a} }.
\end{equation}
More generally, the reduction is equal to $\sqrt{(Q_{\text{red}}/Q_a)}$:
\begin{equation}
E_{\text{th}}^{\text{(probe)}} = \sqrt{ \frac{k_B T Q_{\text{red}}}{\epsilon_0 V Q_a} } .
\end{equation}

The axion to photon conversion power scales as $P_a =P_0  [( Q_c  Q_a )/(Q_c+Q_a)] =P_0 Q_{\text{red}}$, with $P_0 = 10^{-23}\,{\rm W} / 10^4 $.
The axion-induced field becomes:
\begin{equation}
E_a = \sqrt{ \frac{P_0 Q_{\text{red}} Q_c}{\epsilon_0 V \omega} }.
\end{equation}

The resulting SNR is:
\begin{equation}
\mathrm{SNR}(t) = \frac{E_a}{E_{\text{th}}^{\text{(probe)}}} \cdot \sqrt{ \frac{t}{\tau_a} } = \sqrt{ \frac{P_0 Q_c t }{2 \pi k_B T } }.
\label{Eq:SNR1}
\end{equation}
The integrated SNR over time $t$ is proportional to $\sqrt{Q_c t / T}$ and independent of the axion $Q_a$.
The more general equation is given by:
\begin{equation}
\mathrm{SNR}(t) = \frac{E_a}{E_{\text{th}}^{\text{(probe)}}} \cdot \sqrt{ \frac{t}{\tau_a} } = \sqrt{ \frac{P_0 Q_{\text{red}} t }{2 \pi k_B T } }.
\label{Eq:SNR_gen}
\end{equation}
which reduces to Eq.~\ref{Eq:SNR1} above when the cavity quality factor $Q_c \ll Q_a$.

\subsection*{B4. The Effect of Cavity Narrowness on SNR in the High Quality Factor Regime}

When $Q_c \gg Q_a$, the cavity is narrower than the axion linewidth. The axion power saturates:
\begin{equation}
P_a \approx P_0 Q_a,
\end{equation}
and the axion induced electric field amplitude becomes:
\begin{equation}
E_a = \sqrt{ \frac{P_0 Q_a Q_c}{\epsilon_0 V \omega} },
\end{equation}
while the thermal noise field:
\begin{equation}
E_{\text{th}} = \sqrt{ \frac{k_B T}{\epsilon_0 V} }.
\end{equation}

Using $\tau_{\text{coherence}} = (Q_c + Q_a)  /\nu \approx Q_c  /\nu$, the SNR becomes:
\begin{equation}
\mathrm{SNR}(t) = \frac{E_a}{E_{\text{th}}} \cdot \sqrt{ \frac{t}{\tau_{\text{coherence}}} } = \sqrt{ \frac{P_0 Q_a t }{2 \pi k_B T} },
\end{equation}
which can also result from Eq.~\ref{Eq:SNR_gen} above for $Q_c \gg Q_a$.
The integrated SNR is now proportional to $\sqrt{Q_a t/T}$ and independent of the cavity $Q_c$. The SNR scaling as $\sqrt{Q_c}$ holds only when $Q_c \lesssim Q_a$. For $Q_c \gg Q_a$, two competing effects emerge: while the cavity field amplitude continues to increase with $Q_c$, so does the thermal noise field, since the narrower cavity mode enhances thermal fluctuations per unit bandwidth despite admitting less total bandwidth. Simultaneously, only a fraction $Q_a / Q_c$ of the axion power spectrally overlaps with the cavity mode. These effects combine to cancel any further SNR gain, leading to saturation as $Q_c$ increases. The optimal regime is therefore $Q_c \approx Q_a$, where both the axion power and the thermal noise are spectrally matched. Future advances in single-photon detectors will be critical for pushing axion sensitivity further, particularly at low temperatures since they do not suffer from  limitations due to zero point fluctuations~\cite{article:SPC_QLA, article:SPD, article:Kuzmin2020, article:Gatti2023,article:SPC_QLA,article:Kuzmin2020,article:Mottonen2016,article:CARRACK, article:Ahn23,pankratov2025_CASH}.

\subsection*{B5. General SNR Expression Including Quantum Fluctuations}

Using Eq.~(\ref{eq:thermE}) and a procedure similarly as above, the thermal noise within the relevant band width is:
\begin{equation}
E_{\text{th}}^{\text{(probe)}} = \sqrt{ \frac{ \hbar \omega \coth \left( \frac  {\hbar \omega}{2 k_B T} \right ) Q_{\text{red}}}{2 \epsilon_0 V Q_a} },
\end{equation}

This allows one to define the SNR  as:
\begin{equation}
\mathrm{SNR}(t) = \frac{E_a}{E_{\text{th}}^{\rm probe}} \cdot \sqrt{ \frac{t}{\tau_{\text{coherence}}} }=\sqrt{\frac{ P_0 Q_{\text{red}}t}{ \pi \hbar \omega \coth(\frac{\hbar \omega}{2 k_B T})}},
\label{Eq:general}
\end{equation}
which is the most general equation valid in both the classical ($k_B T \gg \hbar \omega$) and quantum ($k_B T \ll \hbar \omega$) regimes and all values of the cavity quality factors. For completeness, in the quantum limit, the SNR becomes:
\begin{equation}
\mathrm{SNR}(t) = \frac{E_a}{E_{\text{th}}} \cdot \sqrt{ \frac{t}{\tau_{\text{coherence}}} } = \sqrt{ \frac{ P_0 Q_{\text{red}} t }{ \pi \hbar \omega} }.
\end{equation}

Thus, by using probing to reduce thermal noise and understanding the interplay with vacuum fluctuations, one can access high-SNR regimes even when the cavity quality factor is not ideally matched to the axion coherence.

\begin{figure*}[t]
  \centering
    \includegraphics[width=\textwidth]{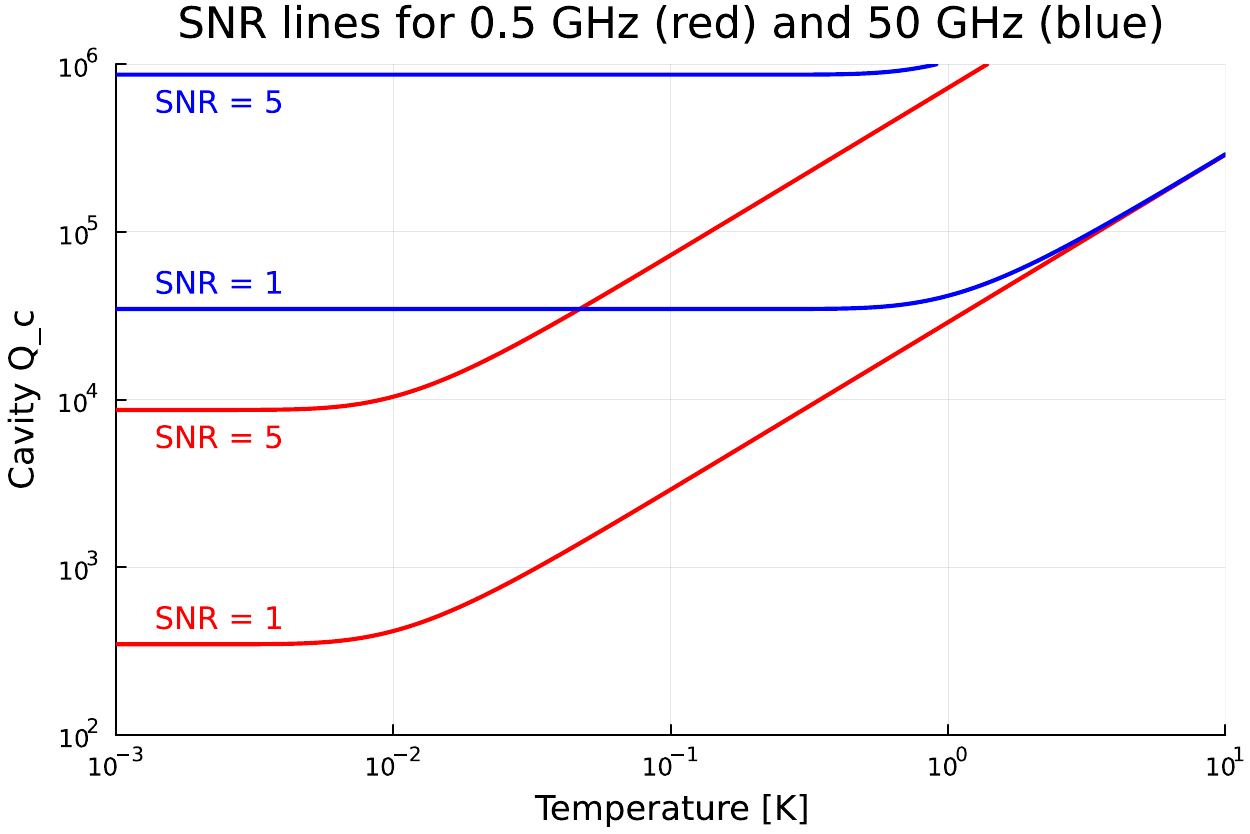}
    \caption{Signal-to-noise ratio   (SNR) as a function of $Q_c$ and $T$ at different operating frequencies. The transition from the quantum regime to the classical regime becomes apparent around 24\,mK for 0.5\,GHz (red), and 2.4\,K for 50\,GHz (blue),  where the thermal photon occupation number begins to exceed the vacuum (zero-point) contribution. The cavity volume is kept constant  at 3.7 liters. The axion to photon conversion power is kept at $10^{-23}$\,W  assuming $Q_c=10^4$ and scaled appropriately for different cavity quality factor values; see text.}
  \label{fig:snr_plots}
\end{figure*}

 The SNR scales as $1/\sqrt{T}$ and benefits  from $\sqrt{Q_c}$ up to the axion quality factor $Q_a$. In the quantum regime, SNR is leveled off at low temperatures and it improves with increasing $\sqrt{Q_c}$, again up to the axion quality factor.
 Figure~\ref{fig:snr_plots} shows that it is possible to overcome the thermal noise background by employing a combination of low temperature and high cavity quality factor.

\section{Bandwidth Considerations via RF Probing in EO Axion Detection}
\label{app:C}

In the electro-optic (EO) probing method for axion dark matter detection, the injection of a coherent radiofrequency (RF) tone into the cavity radically alters the nature of the signal readout. Most importantly, it enables a dramatic reduction of the effective detection bandwidth, directly improving the signal-to-noise ratio (SNR) and enabling sensitivity beyond the conventional quantum limit associated with microwave amplifiers. However, the zero-point fluctuations of the electromagnetic field (vacuum noise) inside the cavity still remain.

\subsection*{C1. Principle of RF Probing}

In traditional haloscope experiments, the axion-induced signal appears as a weak oscillating electric field with a narrow intrinsic bandwidth determined by the velocity dispersion of the dark matter halo:
\begin{equation}
\Delta \nu_a = \frac{\nu_a}{Q_a} \sim 10^{-6} \nu_a \sim 10\,\text{kHz},
\end{equation}
for axion frequencies $\sim 10$\,GHz.

Without RF probing, one must integrate over this full axion bandwidth to recover the signal power, with noise accumulating over the same bandwidth.

By injecting a coherent RF tone $E_p \cos{(\omega_p t)}$ near the expected axion frequency $\omega_a$, the axion field $E_a \cos{(\omega_a t +\phi(t))}$ interferes with it to generate a beat signal at the difference frequency:
\begin{equation}
\omega_b = |\omega_a - \omega_p| \sim \text{few kHz},
\end{equation}

This beat modulates the birefringence of the EO crystal, and appears as a low-frequency modulation in the polarization of a transmitted laser beam.

\subsection*{C2. Detection Bandwidth from Integration Time}

The probing technique concentrates the axion-induced signal into a narrow spectral line. The detection bandwidth is not set by $\Delta \nu_a$, but by the inverse of the integration time:
\begin{equation}
\Delta \nu_{\text{det}} = \frac{1}{t}.
\end{equation}
For $t=3$\,s, $\Delta \nu_{\rm det} \sim 0.33$\, Hz.  This allows one to measure a variance signal in a very narrow frequency bin, thereby reducing noise contribution.

\subsection*{C3. Vacuum Noise Is Present and Probed}

Vacuum fluctuations in the cavity, with RMS field:
\begin{equation}
E_{\text{vac}} = \sqrt{\frac{\hbar \omega}{2 \epsilon_0 V}},
\end{equation}
are broadband and random-phase. When a coherent probe field is injected, vacuum noise beats with it just like the axion field does:
\begin{equation}
E(t) = E_{\text{probe}} \cos(\omega t) + \delta E(t),
\end{equation}
with $\delta E(t)$ including vacuum, thermal, and axion components.

The ellipticity is proportional to:
\begin{equation}
\psi(t) \propto E(t)^2 \propto 2 E_{\text{probe}} \delta E(t),
\end{equation}
which includes vacuum-induced and axion-induced modulation.

Therefore the EO method does \emph{not eliminate} vacuum noise — it still contributes to the measured ellipticity fluctuations.

\subsection*{C4. Why CARAMEL Can Still Resolve the Axion}

What CARAMEL does is leverage:
\begin{itemize}
\item Coherent RF probing to upconvert the axion signal into a narrow beat note;
\item Fabry--Pérot enhancement to increase signal ellipticity;
\item Narrowband optical readout $\Delta \nu_{\rm det} \ll  \Delta \nu_a \le \Delta \nu_{\rm cav}$ to suppress broadband noise.
\end{itemize}

Therefore, although the axion field is weaker than the vacuum field (e.g. $E_a / E_{\rm vac} \sim 10^{-2} - 10^{-1}$) it is concentrated in a single narrow bin, while the vacuum noise is spread across $\Delta \nu \sim \nu / Q_c$. The effective vacuum noise in the detection bin is:
\begin{equation}
E_{\text{vac,eff}} = E_{\text{vac}} \cdot \sqrt{\frac{\Delta f_{\text{det}}}{\Delta f_{\text{cavity}}}},
\end{equation}
which can be $\sim 10^2 - 10^3$ times smaller than the total RMS vacuum field.

\section{Feasibility of the Quantum Readout System (QRS)}
\label{app:D}

\begin{figure*}[t]
  \centering
  \includegraphics[width=0.95\textwidth]{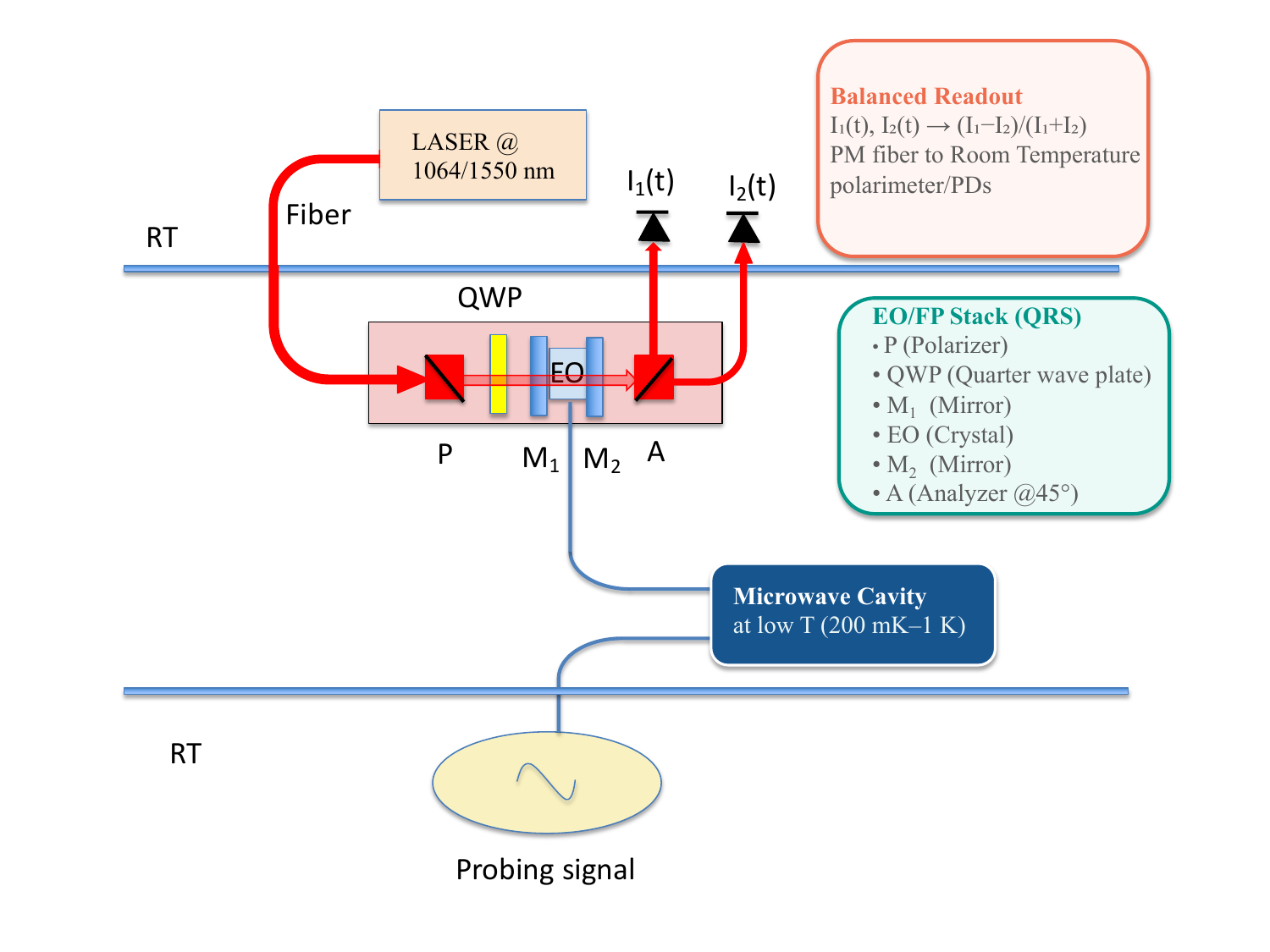}
  \caption{Schematic of the quantum readout system (QRS) used in the feasibility analysis. The microwave cavity is placed at the lowest required operating temperature (typically $0.2$–$1~\mathrm{K}$ depending on the target frequency), while the QRS is mounted on the top plate and coupled to the cavity via a short, impedance-matched RF cable. The electro-optic (EO) crystal is embedded between two high-reflectivity mirrors that form the Fabry–Pérot resonator and is probed by a linearly polarized laser. A polarizer, quarter-wave plate (QWP), and analyzer (oriented at $45^\circ$ relative to the polarizer) convert the axion-induced ellipticity modulation into a detectable polarization rotation. The resulting signal is read out using balanced photodiodes (PDs) at room temperature (RT), which measure the normalized differential output $(I_1(t)-I_2(t))/(I_1(t)+I_2(t))$. Since the required cavity temperatures do not need to fall below $0.2$–$1~\mathrm{K}$ across the targeted frequency range, the laser-induced absorption in the EO crystal must remain below $\sim10\% $ of the available cooling power at that temperature. Depending on the available laboratory infrastructure, the EO/FP stage may alternatively be positioned at the $4~\mathrm{K}$ plate, remaining electrically coupled to the cavity while being only weakly thermally anchored to it.}
  \label{fig:qrs_schematic}
\end{figure*}

\noindent
This appendix summarizes feasibility in three parts: (i) thermal loading {\it vs.}\ QRS location, (ii) sensitivity and demodulation (showing explicitly why no “wash-out” occurs), and (iii) Electro-optic (EO)-crystal length and probing-power scaling with axion frequency. Throughout we reference the schematic in Fig.~\ref{fig:qrs_schematic}.

The key parameters of the EO/FP (QRS) readout have been verified to be mutually consistent across the expected operating range.  
At 10\,GHz the signal from the microwave cavity is mixed down to $\sim10~\mathrm{kHz}$, while at 50\,GHz it is demodulated to $\sim50~\mathrm{kHz}$; the 10-50\,kHz frequencies are orders of magnitude smaller than any relevant optical bandwidth in the Fabry–Pérot cavity, ensuring that the polarization modulation is not averaged out.  

Because the optical absorption coefficient of LiNbO$_3$ is too large to sustain milliwatt-scale laser powers at high finesse, the allowable laser intensity inside the dilution refrigerator is set by the cryogenic thermal budget rather than by optical considerations. For ${\cal F} = 10^{4}$ and a 3~mm EO crystal, the measured absorption corresponds to an absorbed power $P_{\rm abs} \simeq 0.13\,P_\ell$, where $P_\ell$ is the incident laser power. Imposing a conservative thermal limit $P_{\rm tot} = P_{\rm abs} + P_{\rm RF} \le 5~\mu$W and allowing up to $P_{\rm RF} = 2~\mu$W of coherent probe power yields $P_\ell \lesssim 23~\mu$W. This operating point restores the full ${\cal F} = 10^{4}$ FP enhancement while remaining safely within the
cryogenic load budget. At such low optical powers, laser
shot noise is negligible compared to the variance feature
generated by axion–probe interference.

The FP resonator, filled by a LiNbO$_3$ crystal of refractive index $n\simeq2.2$ at 1064 nm, has a free-spectral range (FSR)
\[
\mathrm{FSR}=\frac{c}{2nL_{\mathrm{FP}}}, \qquad
\Delta\nu_{\mathrm{FP}}=\frac{\mathrm{FSR}}{\cal{F}}. \tag{D1}
\]
For a  $L_{\mathrm{FP}}=3~\mathrm{mm}$ (assumed equal to the crystal length $L_c$) and ${\cal F}=10^4$, one obtains 
$\mathrm{FSR}\!\approx\!22.7~\mathrm{GHz}$ and a linewidth 
$\Delta\nu_{\mathrm{FP}}\!=\!2.3$$~\mathrm{MHz}$, 
comfortably larger than the modulation frequencies (10–50 kHz).  
The EO modulation therefore lies fully within the same FP resonance, guaranteeing faithful transmission of the sidebands and preserving the ellipticity-variance measurement.

To maintain a large enough signal-to-noise ratio as the target frequency increases, the EO-crystal length can be reduced in proportion to the axion frequency (e.g., $L_c:3~\mathrm{mm}\!\rightarrow\!0.6~\mathrm{mm}$ for 10 → 50 GHz) while the microwave probe power $P_{\mathrm{RF}}$ can be raised to compensate for it.  
The shorter crystal reduces optical absorption and allows higher circulating optical power, while the higher $P_{\mathrm{RF}}$ compensates for the smaller EO-interaction volume.  
Under these conditions the demodulated signal bandwidth (10–50 kHz) remains far below $\Delta\nu_{\mathrm{FP}}$, confirming that all parameter combinations discussed are thermally and dynamically feasible.
\subsection*{D1. Thermal Location and Loading}

The thermal feasibility of the QRS depends primarily on the absorbed optical power,
\[
P_{\mathrm{abs}} = \alpha\,L_c\,\frac{{\cal F}}{\pi}\,P_\ell,
\tag{D2}
\]
where $\alpha$ is the absorption coefficient of the EO crystal, $L_c$ its optical length, ${\cal F}$ the finesse of the FP cavity, and $P_\ell$ the incident laser power.  
For LiNbO$_3$ at 1064~nm, $\alpha\simeq(3\text{–}5)\times10^{-3}~\mathrm{cm^{-1}}$ at room temperature and decreases by roughly one order of magnitude at cryogenic temperatures.  

Using $L_c=3~\mathrm{mm}$, a finesse of $10^4 $, and $P_\ell=0.1~\mathrm{mW}$ gives
\[
P_{\mathrm{abs}}\approx 
\left(4\times10^{-3} \times 0.3\right)\frac{\cal{F}}{\pi}P_\ell
  \simeq 0.13\times P_\ell
  \approx 13~\mu\mathrm{W},
\tag{D3}
\]
which remains well below the available cooling power of typical dilution refrigerators at 0.2-1\,K.  
Had the original $P_\ell=10$~mW been retained, the absorbed power would increase by two orders of magnitude, clearly exceeding the 0.2-K cooling capability but not at 1\,K.  
Thus, lowering the laser power to $P_\ell\!\sim\!0.1$~mW alone suffices to make the system thermally viable while keeping the SNR very high.

Consequently, the QRS can safely be located either directly on top of the cavity—thermalized at the same stage as the resonator—or at the 4-K plate, depending on the target axion frequency.  
For operation below $\sim15~\mathrm{GHz}$, the cavity typically resides near 0.1–0.2~K and the QRS can be co-mounted on its top plate without imposing a thermal load greater than $\leq 10\%$ of the available cooling power.  
At higher frequencies (e.g., 50~GHz), the signal-to-noise ratio saturates already near 1~K, where the available cooling power is larger by $T^2$ more than one order of magnitude, permitting optical powers up to the milliwatt level without penalty.  

The microwave coupling between the cavity and QRS is realized through a short, thermally anchored coaxial line (typically $\sim10~\mathrm{cm}$).  
Since the injected RF power is only in the range of 200~nW–1~$\mu$W, the corresponding Joule dissipation in the coupling network remains negligible compared with the cooling power at all stages.
\subsection*{D2. Demodulated (Variance) Sensitivity}

In the probing scheme, the EO/FP stage converts the axion-driven cavity field into a polarization signal whose
\emph{mean} is time independent, while its \emph{variance} fluctuates on the axion coherence timescale.
The relevant frequency scale is set solely by the axion coherence bandwidth
\[
\Delta \nu_a \simeq \nu_a (v/c)^2,
\]
with $v\approx 10^{-3} c$, which gives $\Delta \nu_a \sim 10~\mathrm{kHz}$ at $\nu_a=10~\mathrm{GHz}$ and $\sim 50~\mathrm{kHz}$ at $\nu_a=50~\mathrm{GHz}$.
The power spectral density of the ellipticity (or differential intensity) therefore exhibits a narrow feature of width
$\Delta \nu_a$ around baseband. We sample at $\nu_s \gg 2\,\Delta \nu_a$ and estimate the variance within a window
$B \sim \Delta \nu_a$ (or, equivalently, $B \sim 1/t$ for matched integration).

The signal-to-noise ratio (SNR) of the variance estimate follows the radiometer law for power detection,
\[
\mathrm{SNR}=\frac{P_s}{k_B T_n}\sqrt{\frac{t}{B}},
\]
where $P_s$ is the power in the variance feature, $T_n$ the equivalent noise temperature,
$t$ the integration time, and $B$ the analysis bandwidth.
For $t \le \tau_c$ (with $\tau_c \sim 1/\pi \Delta \nu_a$), matched filtering implies $B \sim 1/t$ and hence
$\mathrm{SNR} \propto t$.
For $t \gg \tau_c$, the data contain $N=t/\tau_c$ statistically independent coherence segments,
and the optimal estimator’s variance decreases as $1/\sqrt{N}$, yielding the familiar
\[
\mathrm{SNR} \propto \sqrt{t/\tau_c}.
\]
This is the same long-time $t^{1/2}$ scaling used in phase-preserving haloscope analyses; here it applies to the
variance feature centered at baseband with width $\Delta f_a$.
\subsection*{D3. Frequency–Crystal–Power Scaling}

Because the microwave cavity and its stored field remain constant, maintaining the same SNR across the 10–50~GHz range requires only modest adjustments in the EO parameters.  
Shortening the EO crystal at higher frequencies increases the FSR.  
At the same time, the microwave probe power can be increased within the same cavity coupling configuration to compensate for the smaller EO interaction length, preserving the effective modulation depth.
Because the absorbed optical load remains fixed by a
$\le 5~\mu$W budget at the low frequencies (around 1\,GHz), the allowable laser power scales weakly
with $L_c$, and the system maintains nearly constant SNR
($\mathrm{SNR} \approx 38$ at $t = 3$~s) over the full frequency span.
Representative parameter choices consistent with the thermal and optical constraints are summarized in Table~\ref{tab:scaling}. The parameter values listed on Table~\ref{tab:benchmark} are feasible as long as the QRS is mounted at the 4\,K plate of the dilution refrigerator, where the cooling power is adequate.
\begin{table}[t]
\centering
\caption{Representative QRS parameters as a function of
frequency, assuming fixed cavity volume and a total thermal
load $\le 5~\mu$W. The operating point maintains ${\cal F} = 10^{4}$
across the full band.}
\begin{tabular}{lccc}
\hline\hline
Parameter & 10 GHz & 30 GHz & 50 GHz \\
\hline
$T_{\rm cav}$ (K) & 0.2 & 0.5 & 1.0 \\
$L_c$ (mm) & 3.0 & 1.5 & 0.6 \\
${\cal F}$ & $10^{4}$ & $10^{4}$ & $10^{4}$ \\
$P_\ell$ ($\mu$W) & 23 & 23 & 23 \\
$P_{\rm RF}$ ($\mu$W) & 2.0 & 2.0 & 2.0 \\
FSR (GHz) & 22.7 & 68 & 113 \\
$\Delta\nu_{\rm FP}$ (MHz) & 2.27 & 6.8 & 11.3 \\
$f_{\rm mod}$ (kHz) & 10 & 30 & 50 \\
SNR (3 s) & $\sim 38$ & $\sim 38$ & $\sim 38$ \\
\hline\hline
\end{tabular}
\label{tab:scaling}
\end{table}

This scaling shows that as the frequency increases, the EO crystal becomes shorter and optically more efficient, while the broadened FSR and available probe power ensure that both the thermal and dynamic conditions remain easily satisfied.

\noindent
\textbf{Implementation note.} If thermal budget on the coldest stage is tight, one could place the QRS on the 4\,K plate and couple the cavity to the EO capacitor via a short, impedance-matched line with \emph{in-situ} attenuation mounted at the cavity temperature so that the cavity does not see 4\,K noise. The optical readout remains unchanged; the demodulated signal at \(\sim\)10~kHz is unaffected, and the FP still satisfies \(f_\mathrm{mod}\ll \Delta\nu_\mathrm{FP}\).

\bibliography{main}

\end{document}